\documentclass{eas}
\usepackage{astron}
\usepackage{amsmath,amssymb}
\usepackage{graphicx}
\usepackage{multirow}
\usepackage[svgnames]{xcolor}
\def\msol{\ensuremath{M_\odot}}
\def\lsol{\ensuremath{L_\odot}}
\def\rsol{\ensuremath{R_\odot}}
\def\zsol{\ensuremath{Z_\odot}}
\def\teff{\ensuremath{T_\text{eff}}}
\def\ooc{\ensuremath{\Omega/\Omega_\text{crit}}}
\def\vvc{\ensuremath{V/V_\text{crit}}}
\begin{document}

%%-----------------------------
%%      the top matter
%%-----------------------------
\title{Red supergiants and stellar evolution}
\runningtitle{RSG evolution}
\author{Sylvia Ekstr\"om}\address{Geneva Observatory, University of Geneva, Maillettes 51 - Sauverny, CH - 1290 Versoix, Switzerland}
\author{Cyril Georgy}\address{CRAL, ENS-Lyon, 46 All\'ee d'Italie, 69369 Lyon C\'edex 07, France}
\author{Georges Meynet}\sameaddress{1}
\author{Jos\'e Groh}\sameaddress{1}
\author{Anah\'i Granada}\sameaddress{1}
\begin{abstract}
We review the significant role played by red supergiants (RSGs) in stellar populations, and some challenges and questions they raise for theoretical stellar evolution. We present how metallicity and rotation modify the way stars go to the red part of the Hertzsprung-Russell diagram or come back from it, and how RSGs might keep a trace of their main-sequence evolution. We compare theoretical population ratios with observed ones. 
\end{abstract}
\maketitle
%%-----------------------------
%%      your text
%%-----------------------------
%===================================================================================
\section{Introduction}
%------------------------------------------------------------------------------------------------------------------------------------------------
\subsection{Why red supergiants are interesting objects}
If we make the hypothesis that all stars with a mass between $8$ and $25\,\msol$ go through a red-supergiant (RSG) phase, it means that 80\% of all massive stars will be a RSG one day. This phase is a crucial one for the mass-loss history of the star (see C. Georgy's contribution in this conference and references therein). RSG are also supposed to be the direct progenitors of the most numerous type of core-collapse supernovae (cc-SNe): the type II.

We present an evolutionary scheme for massive stars in Table~\ref{tabConti}, inspired from Conti \cite*{Conti1975a}, based on solar-metallicity non-rotating models.
\begin{table}
\scalebox{.86}{%
\begin{tabular}{|lllc|}
\hline
$M > 60\,\msol$: & {\small O $\rightarrow$ Of/WNL $\rightarrow$ LBV $\rightarrow$ WNL  $\rightarrow$ (WNE) $\rightarrow$ WC}& $\rightarrow$ SN Ibc & \multirow{4}{*}{\large \color{Grey}{WR}}\\
 & & & \\
{$M = 40-60\,\msol$}: & {\small O $\rightarrow$ BSG $\rightarrow$ LBV $\rightarrow$ WNL $\rightarrow$ (WNE) $\rightarrow$ WC} & $\rightarrow$ SN Ibc & \\
% & & & \\
\hline
\multicolumn{1}{l}{\null} & & & \multicolumn{1}{c}{\null} \\
\multicolumn{1}{l}{$M = 30-40\,\msol$:} & {\small O $\rightarrow$ BSG $\rightarrow$ RSG $\rightarrow$ WNE$\rightarrow$ WCE} & $\rightarrow$ SN Ibc & \multicolumn{1}{c}{\null} \\
\multicolumn{1}{l}{\null} & & & \multicolumn{1}{c}{\null} \\
\hline
% & & & \\
{$M = 25-30\,\msol$}: & {\small O $\rightarrow$ (BSG) $\rightarrow$ RSG $\rightarrow$ (YSG?)} & $\rightarrow$ SN II-L/b & \multirow{3}{*}{\large \color{Grey}{RSG}}\\
 & & & \\
{$M = 10-25\,\msol$}: & {\small O $\rightarrow$ RSG $\rightarrow$ (Ceph. loop for $M < 15\,\msol$) $\rightarrow$ RSG} & $\rightarrow$ SN II-P & \\
% & & & \\
\hline
\end{tabular}
}
\caption{Modified Conti scenario for the evolutionary scheme of massive stars. \label{tabConti}}
\end{table}
In this scheme, all massive stars with $M\lesssim30\,\msol$ end their life as a RSG, while all massive stars with $M\gtrsim40\,\msol$ end their life as Wolf-Rayet (WR) stars. Between $30$ and $40\,\msol$, the mass loss during the RSG phase might be sufficient to strip the external layers deeply enough for the star to become a WR star. It must be clearly stated here that the real evolutionary path between the different stellar types depends on the physics considered in the models, and might be significantly modified by the mass-loss rates used, the overshooting amplitude, the inclusion or not of the effects of rotation and of magnetic fields. A large fraction of massive stars might also undergo interactions in close binaries \cite{Sana2012a}, and thus follow a quite different evolutionary path.

RSGs are easily seen in external galaxies \cite{Lancon2009a,Drout2012a,Neugent2012a} which makes them also interesting objects because they can be used as metallicity or distance indicators. Davies \etal\ \cite*{Davies2010a} and Bergemann \etal\ \cite*{Bergemann2012a} proposed the RSGs as metallicity indicators, showing that the determination of $Z$ is robust even when only modest resolution ($R=1-3000$) is available. They are proposed as distance indicators (see B. Davies' contribution in this proceedings). Since the RSG phase is limited in mass range and is a relatively short  advanced stage of stellar evolution ($\leq 10\%$ of the lifetime), they are also proposed as age indicators \cite{Lancon2009a}.

%------------------------------------------------------------------------------------------------------------------------------------------------
\subsection{Some present-day challenges in the physics of RSGs}
In 1D stellar-evolution codes, many mechanisms cannot be derived from {\em ab initio} physical principles, and are thus parametrised. It is the case for a mechanism that is of extreme importance for RSG modelling: convection.

Looking at the structure of a RSG, we see that while the core is encompassed in roughly half a solar radius, the star itself might extend to around $800\,\rsol$, most of it being convective. The modelling of such objects makes it necessary to have a good treatment of convection, which is still out of reach of 1D stellar-evolution codes. This convection is non adiabatic, and supersonic turbulence needs to be taken into account \cite{Maeder2009a}.

Another challenge concerns the mass loss. It is implemented in the codes thanks to prescriptions offered in the literature. However there is a debate about the rates inferred for RSGs. Some observations \cite{Mauron2011a} give support to the de Jager \etal\ \cite*{deJager1988a} rate usually implemented in the codes for this stage of the evolution. Others, on the contrary, challenge these rates and suggest much larger ones \cite{vanLoon2005a}. A large uncertainty comes from the unknown nature of the mass loss in that stage: steady and regular, or outbursting? If the mass loss occurs by bursts \cite{Humphreys2005a,Smith2009a,Moriya2011a}, the probability of observing a star precisely in a burst episode might be quite low, so the mass-loss rates are probably underestimated. In the stellar modelling perspective, we are bound to use steady rates and thus implement a mass-loss recipe that would average over time the total mass-loss episodes gone through by the RSG. However it is unclear how to derive this time-averaged mass loss out of burst episodes, and moreover the steady approximation might modify the global behaviour of the star.

%------------------------------------------------------------------------------------------------------------------------------------------------
\subsection{Some crucial questions for the evolutionists}
The evolution to or from a RSG stage is not straightforward to understand. We need to determine the physical conditions driving a star to be or not to be a RSG through the following questions:
\begin{itemize}
  \item When does a massive star enter into the RSG phase?
  \item How long does this phase last?
  \item What is the impact of the physical processes available in stellar evolution codes (treatment of convection, mass loss, rotation, magnetic fields, multiplicity)?
  \item What types of cc-SNe arise from RSG?
  \begin{itemize}
    \item type II-P? II-L?
    \item Is there a SN event if the core collapse results in a black hole?
  \end{itemize}
\end{itemize}

In Section~\ref{SecRSG}, we shall address the first three questions by studying theoretical models published recently \cite{Ekstrom2012a,Georgy2013a}. In Section~\ref{SecPop}, we shall consider RSGs as a population and compare theoretical ratios to the observed ones.

%===================================================================================
\section{To be or not to be a RSG\label{SecRSG}}
Let us first quickly recall the main physical ingredients of the models (the details can be found in Ekstr\"om \etal\ 2012). Compared to the previous complete grids of Schaller \etal\ \cite*{Schaller1992a}, some reaction rates have been updated, the abundances have been chosen to be those of Asplund \etal\ \cite*{Asplund2005a}, and the opacities have been updated to match the abundances. The effect of rotation are implemented according to Zahn \cite*{Zahn1992a} for the horizontal turbulence diffusion coefficient and to Maeder \cite*{Maeder1997a} for the shear diffusion coefficient.

Mass loss is an important ingredient for the modelling of massive stars. In the models, we use Vink \etal\ \cite*{Vink2001a} for the O-type stars and for the RSGs we use a linear fit based on observations by Crowther \cite*{Crowther2001a} that yields mass-loss rates similar to the de Jager \etal\ \cite*{deJager1988a} prescription. For stars with $M\geq20\,\msol$, we multiply the mass-loss rate by a factor of 3 if the star has some external layers that have a supra-Eddington luminosity.

%------------------------------------------------------------------------------------------------------------------------------------------------
\subsection{Becoming a RSG\label{SecRSGCrossing}}
At solar metallicity ($\zsol=0.014$), our models cross quite rapidly the Hertzsprung-Russell (HR) gap, spending most of the core He-burning phase in the red-supergiant region (see G. Meynet's contribution in this conference for a discussion on the physical mechanisms involved).

\begin{figure}
\includegraphics[width=.48\textwidth]{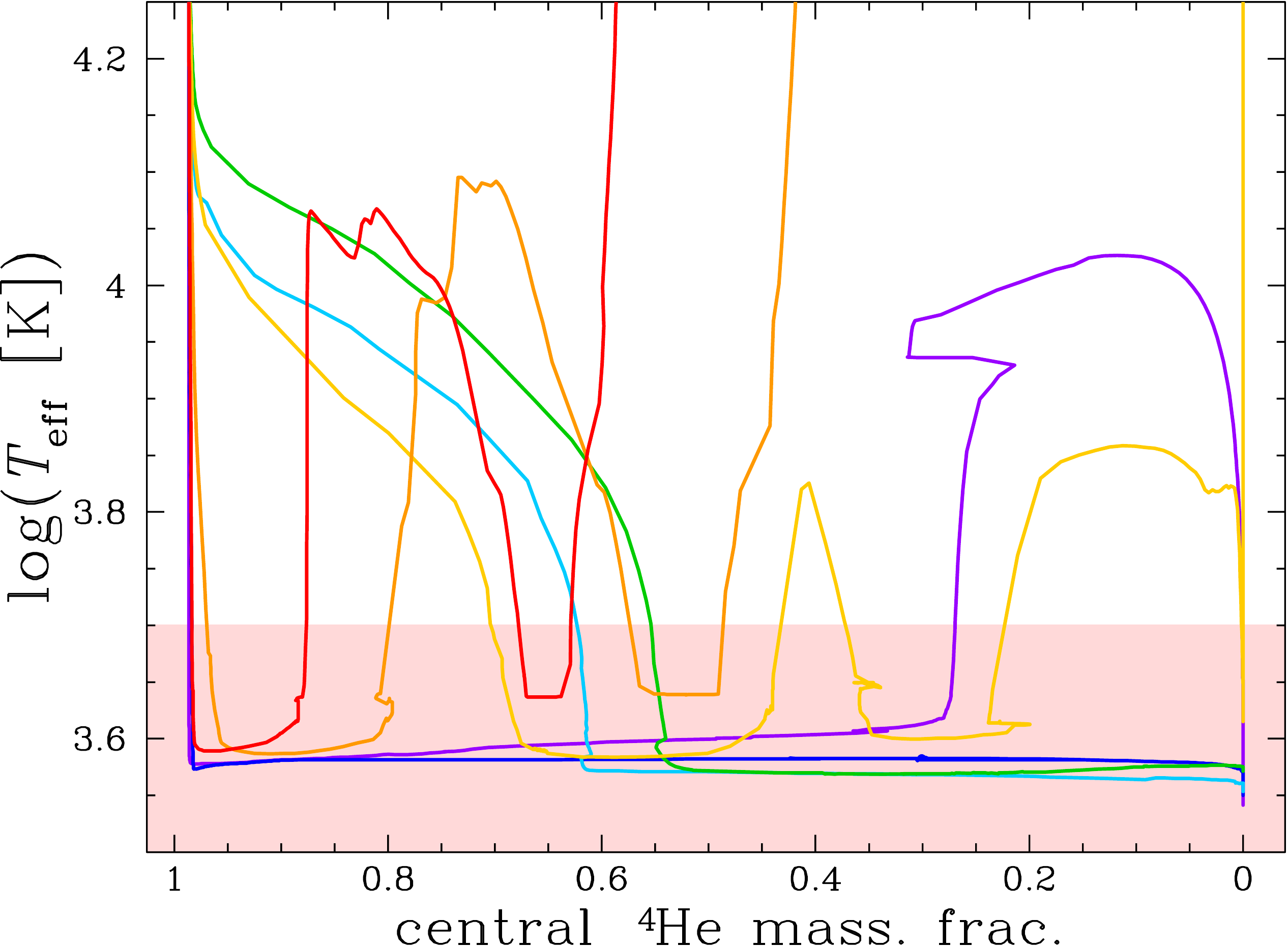}\hfill\includegraphics[width=.48\textwidth]{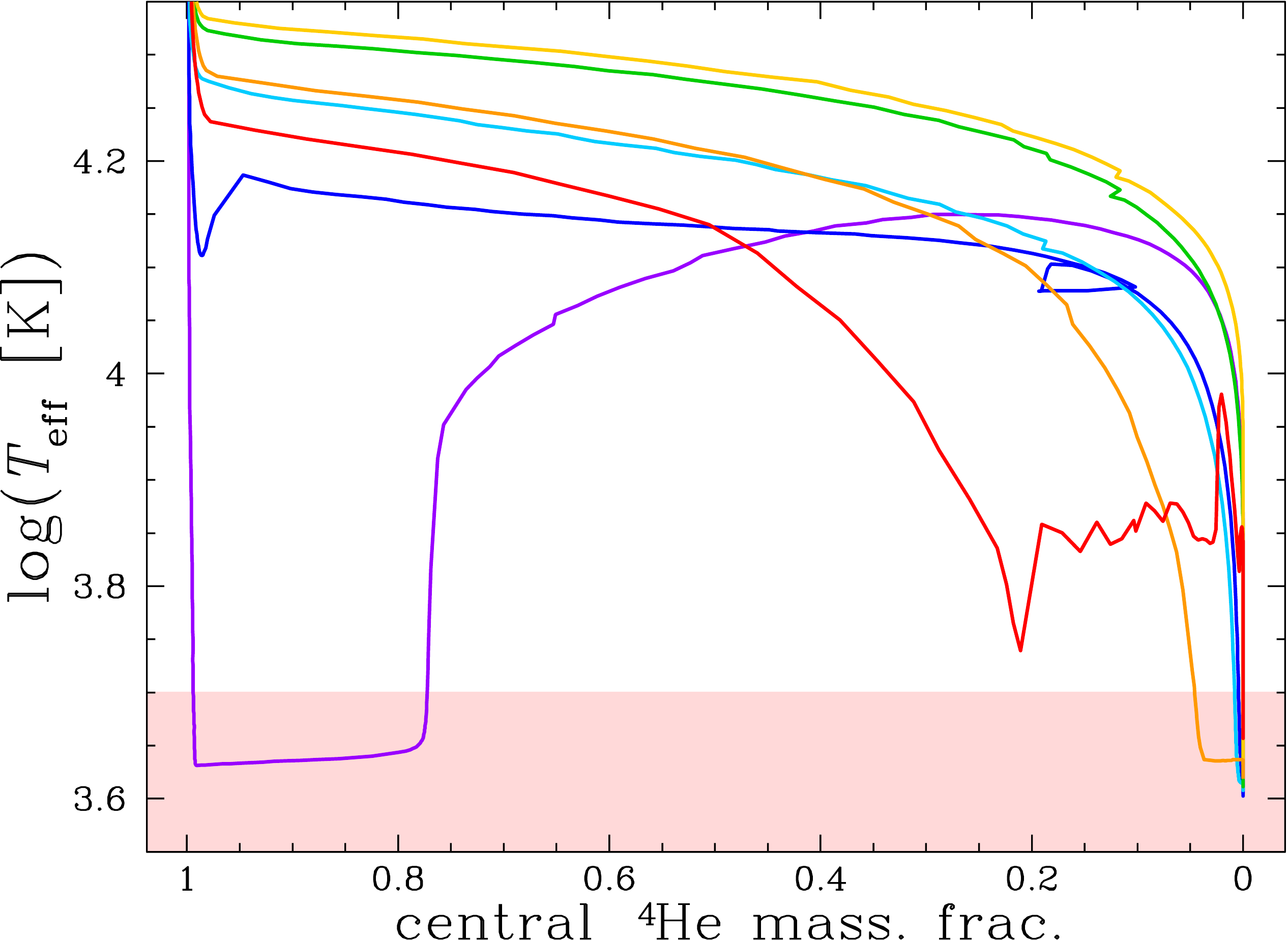}
\caption{Evolution of the \teff\ as a function of the central He combustion (time evolution goes from left to right) in  non-rotating models. The RSG domain is shaded in red. The models of $9\,\msol$ (purple), $12$ (dark blue), $15$ (cyan), $20$ (green), $25$ (yellow), $32$ (orange), and $40\,\msol$ (red) are represented. {\em Left:} solar metallicity ($\zsol=0.014$). {\em Right:} SMC metallicity ($Z=0.002$), models from Georgy \etal\ (in prep).\label{FigYTeffZnr}}
\end{figure}
Figure~\ref{FigYTeffZnr} shows the behaviour of the \teff\ as a function of the central He abundance for two different values of $Z$. The mass range for becoming a RSG ($\log(\teff)\leq3.70$, red-shaded area on the figure) is between $9$ and $40\,\msol$ for our non-rotating models at \zsol\ (left panel), and the associated age range is between $4.5$ and $30$ Myr. It appears that the $9$ and $12\,\msol$ models cross the HR diagram right after the main sequence (MS), before He ignition. The $15$ and $20\,\msol$ models cross the gap before having burnt half of the central He and finish their life in the red. The $25\,\msol$ model crosses the gap slightly earlier but then makes two blue loops during core He burning, and finishes its life in the blue. Both the $32$ and $40\,\msol$ models make a quick incursion into the red, then make a large blue loop, come back very briefly and finish their life as WNE.

As discussed in G. Meynet's contribution in this conference, when the metallicity is lower, the star enters into the RSG stage at a more advanced stage of the core Helium-burning phase (Fig.~\ref{FigYTeffZnr}, {\em right}). This implies that the duration of the RSG phase is much shorter. While at solar metallicity, RSGs are He-burning stars, at low $Z$ they are C-burning stars (see J. Groh's contribution in this conference). We shall see below the consequences it bears on stellar populations.

\begin{figure}
\includegraphics[width=.48\textwidth]{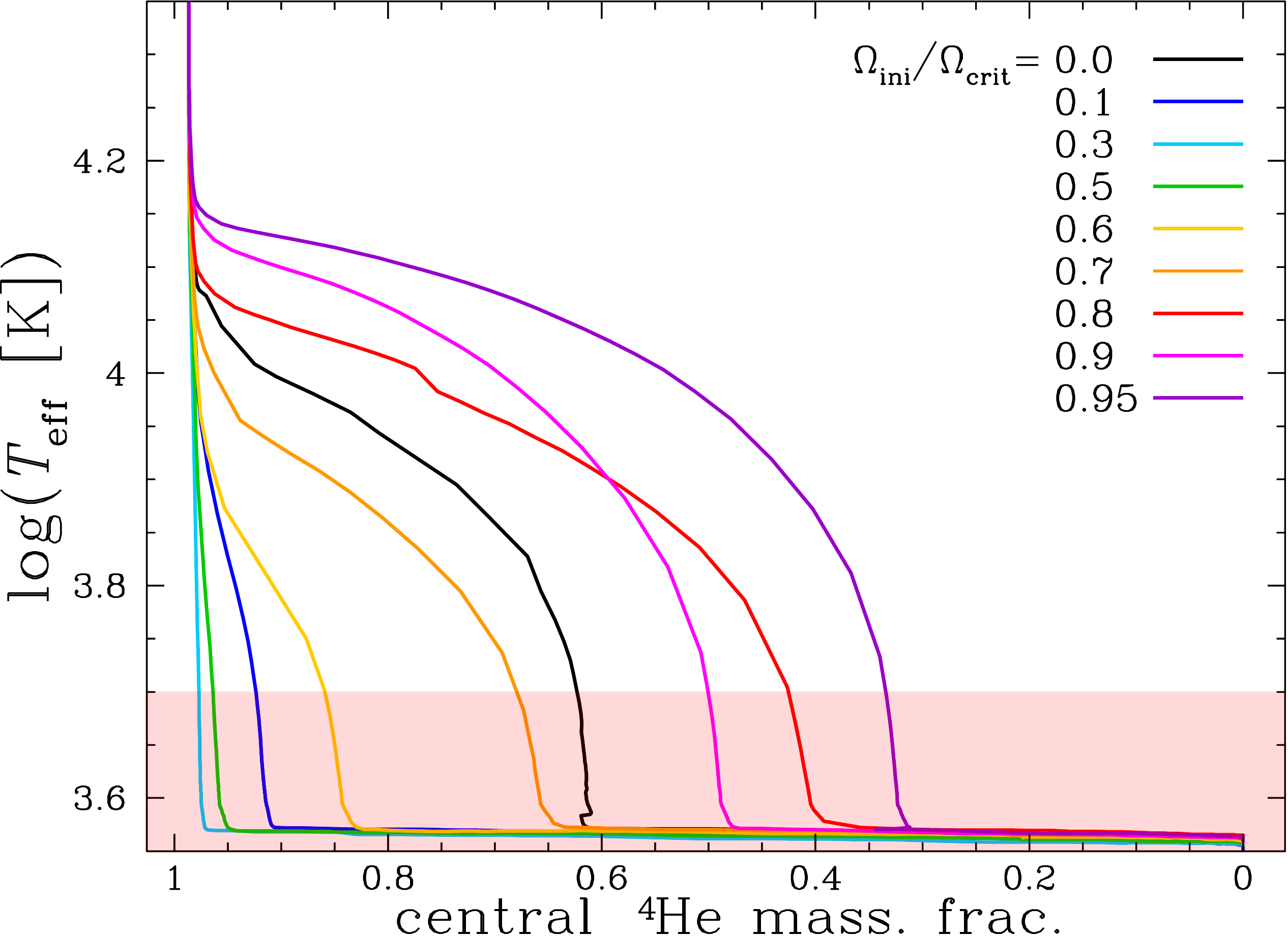}\hfill\includegraphics[width=.48\textwidth]{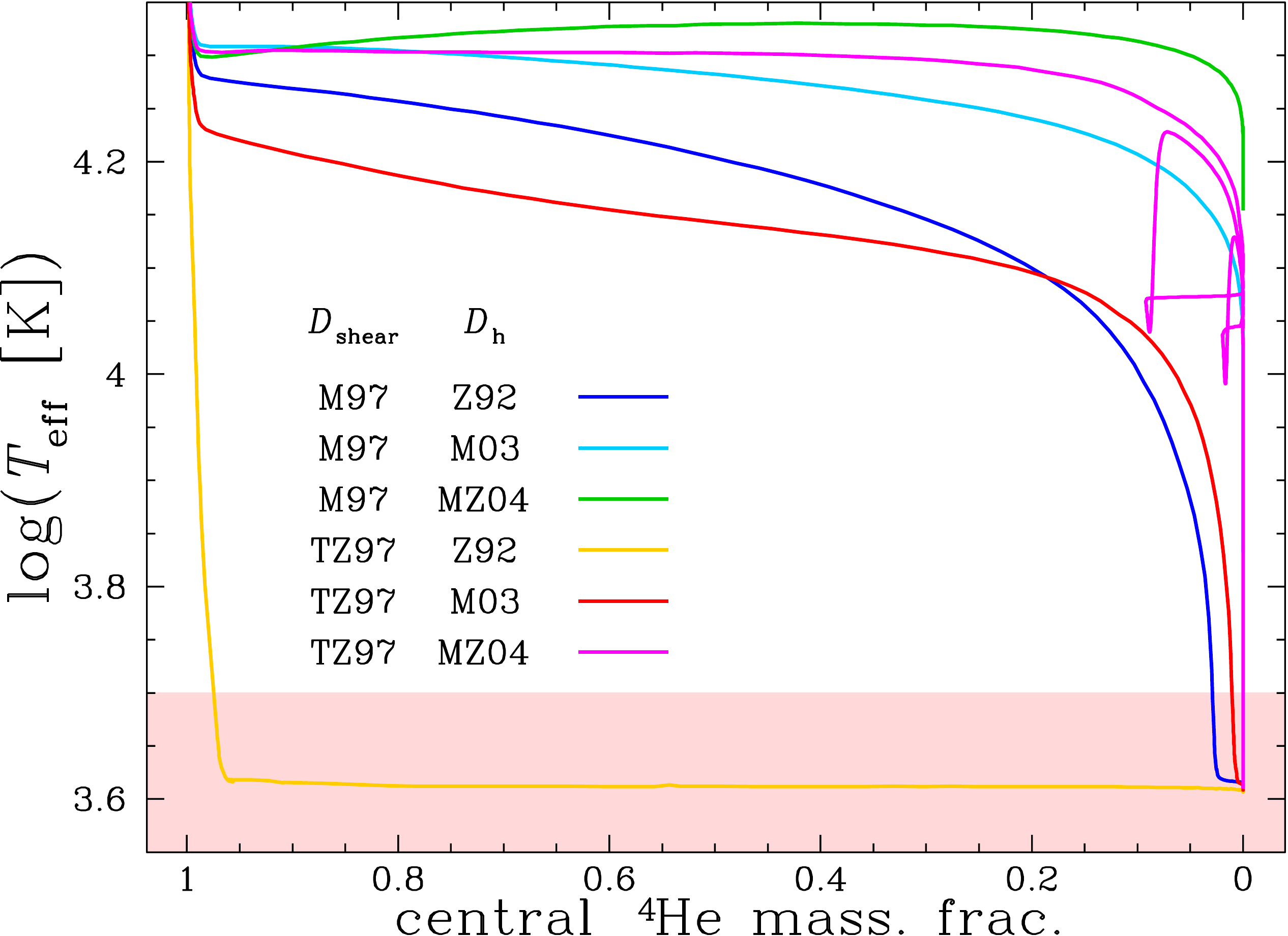}
\caption{Same as Fig.~\ref{FigYTeffZnr}, for $15\,\msol$ models. {\em Left:} solar metallicity ($Z=0.014$), various rotation rates (models from Georgy \etal\ 2013). {\em Right:} SMC metallicity ($Z=0.002$), various rotation prescriptions (models from Meynet \etal\ 2013, with $\ooc=0.50$ initially).\label{FigYTeffrot}}
\end{figure}
Rotation tends to keep the most massive stars in the blue, so the mass range in which we expect to see stars evolve into RSGs is reduced in our $\vvc=0.40$ ($\ooc=0.56$) models: between $9$ and $25\,\msol$ only. The age range associated with this phase is between  $8$ and $35$ Myr. Looking more closely at the effects of rotation, we note that the trend changes depending on the rotation rate. Passing from $\vvc=0$ to $0.20$ ($\ooc=0$ to $0.3$) favours the redward crossing of the HR diagram. But if rotation rates become higher, the trend gets inversed and the crossing occurs at later times during He burning (Fig.~\ref{FigYTeffrot}, {\em left}, models from Georgy \etal\ 2013). In real stellar populations, we expect a velocity distribution, and thus a diversity of behaviours. One must bear in mind that there are several prescriptions for implementing the effects of rotation into the codes. In the literature, we find two different shear-diffusion coefficients \cite{Maeder1997a,Talon1997a} and three different horizontal-turbulence coefficients \cite{Zahn1992a,Maeder2003a,Mathis2004a}. Depending on the set of prescriptions used, the behaviour might be quite different \cite{Meynet2013a}, as illustrated in Fig.~\ref{FigYTeffrot} ({\em right}).
%------------------------------------------------------------------------------------------------------------------------------------------------
\subsection{Back to the blue}
The duration of the RSG phase depends strongly on the mass loss experienced by the star (see C. Georgy's and G. Meynet's contributions, and references therein). According to the observations of Smartt \etal\ \cite*{Smartt2009a}, there are no RSG progenitors for SN II-P above around $18\,\msol$. This coincides with the increased sensitivity to supersonic convection at $20\,\msol$ and above \cite{Maeder2009a}, and thus potentially higher mass-loss rates. In our rotating models at \zsol, the lower-mass stars end their lifes as RSGs and the higher-mass stars move back to the blue. The transition between both behaviours occurs near $18\,\msol$.  In the models, it is most probably due to the treatment of the supra-Eddington layers, however it illustrates well the role played by the mass loss.
%------------------------------------------------------------------------------------------------------------------------------------------------
\subsection{Keeping a trace from the past}
\begin{figure}
\includegraphics[width=.48\textwidth]{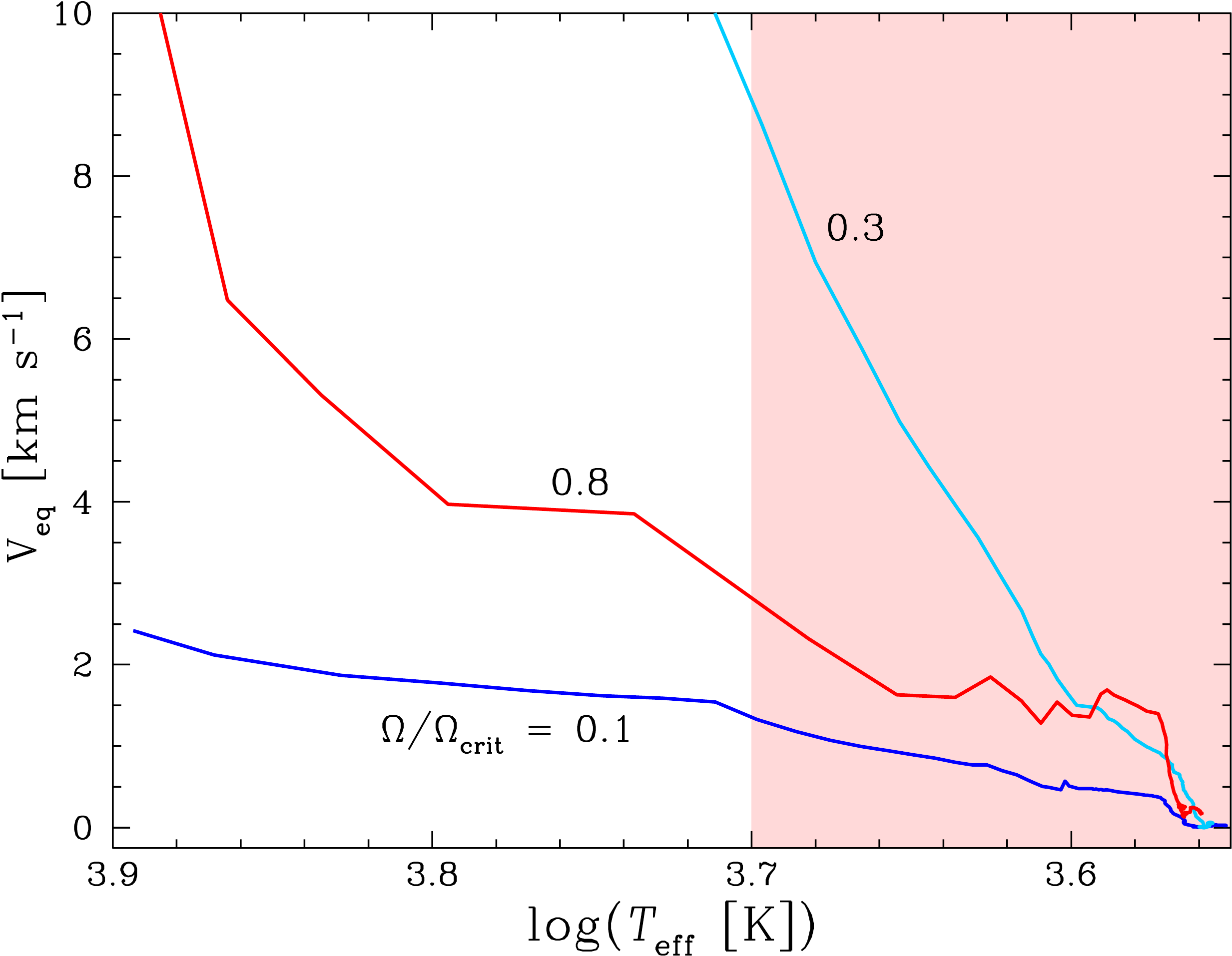}\hfill\includegraphics[width=.48\textwidth]{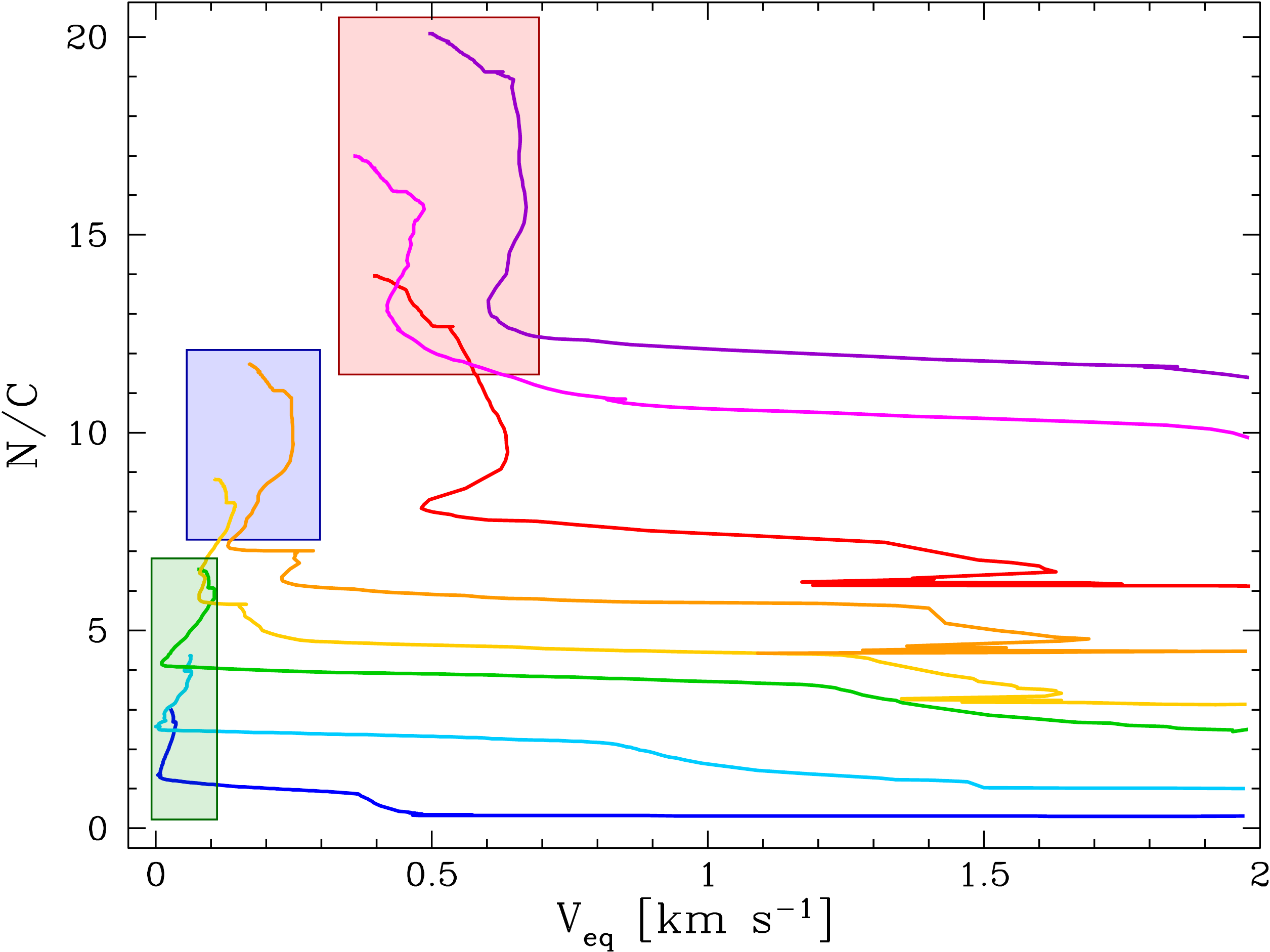}
\caption{$15\,\msol$ models at \zsol, from Georgy \etal\ 2013. {\em Left:} evolution of the surface velocity as a function of the \teff\ during the HR-gap crossing. {\em Right:} N/C surface abundances (in number) as a function of the surface velocity in the RSG phase (same colour coding as Fig.~\ref{FigYTeffrot}, {\em left}).\label{FigVNC}}
\end{figure}

One can wonder whether a rotating star might keep a trace of its former rotation once it has undergone the tremendous inflation leading to the RSG phase. If the crossing of the HR gap is accomplished within a Kelvin-Helmholtz timescale (which is the case for models crossing right after the MS and before any He burning), the surface velocity of the star will evolve according to angular-momentum conservation. On the contrary, if the crossing occurs during part of the He-burning phase (which is the case for the most rapid rotators), the winds have time to remove some angular momentum during the crossing, but also the transport mechanisms have time to bring some angular momentum to the surface, and the evolution of the surface velocity is more complex (Fig.~\ref{FigVNC}, {\em left}).

Once in the red, we could expect the subsequent dredge-up to erase any previous enrichment, but this is not the case. Figure~\ref{FigVNC} ({\em right}) shows that the enrichment of the surface of a former rapidly-rotating star remains more highly enriched (by a factor of 50-60) than that of a non-rotating star. We can distinguish three endpoint zones: one were stars have very low surface velocity ($V_\text{eq}\leq0.1$ km/s) and surface enrichment (N/C $\leq7$, green zone in the figure); one with surface velocities around 0.2 km/s and N/C around 10 (blue zone); and the zone where the former most rapidly-rotating stars end, with velocities around 0.5 km/s and N/C $\gtrsim 15$ (red zone).

%===================================================================================
\section{RSGs as a population\label{SecPop}}
Stellar-population ratios are good indicators to compare theoretical tracks with "real" stars. There is a difference between ratios observed in a coeval population (open cluster) or in a general population with a constant star-formation rate:
\begin{itemize}
  \item in a {\bf coeval population}, the ratio between two types of stellar populations ($A$ and $B$) can be predicted by:
  $$
  \frac{A}{B} = \frac{\int_{M_{A,\text{min}}}^{M_{A,\text{max}}}\,\phi(m)\,\text{d}m}{\int_{M_{B,\text{min}}}^{M_{B,\text{max}}}\,\phi(m)\,\text{d}m}
  $$
  where we just have to know the initial mass range leading to each population $\left[M_{A,\text{min}}\,...\,M_{A,\text{max}}\right]$ and $\left[M_{B,\text{min}}\,...\,M_{B,\text{max}}\right]$ and integrate it over an IMF $\phi(m)$. It thus brings constraints on the {\bf mass limits} inferred;
  \item in a {\bf constant star-formation population}, the same ratio involves also the duration of the phase in which the star is of the given type ($\tau_A$ and $\tau_B$):
  $$
  \frac{A}{B} = \frac{\int_{M_{A,\text{min}}}^{M_{A,\text{max}}}\,\tau_A\,\phi(m)\,\text{d}m}{\int_{M_{B,\text{min}}}^{M_{B,\text{max}}}\,\tau_B\,\phi(m)\,\text{d}m}.
  $$
  It thus brings constraints on the mass limits and the {\bf durations}.
\end{itemize}
\begin{figure}
\includegraphics[width=.48\textwidth]{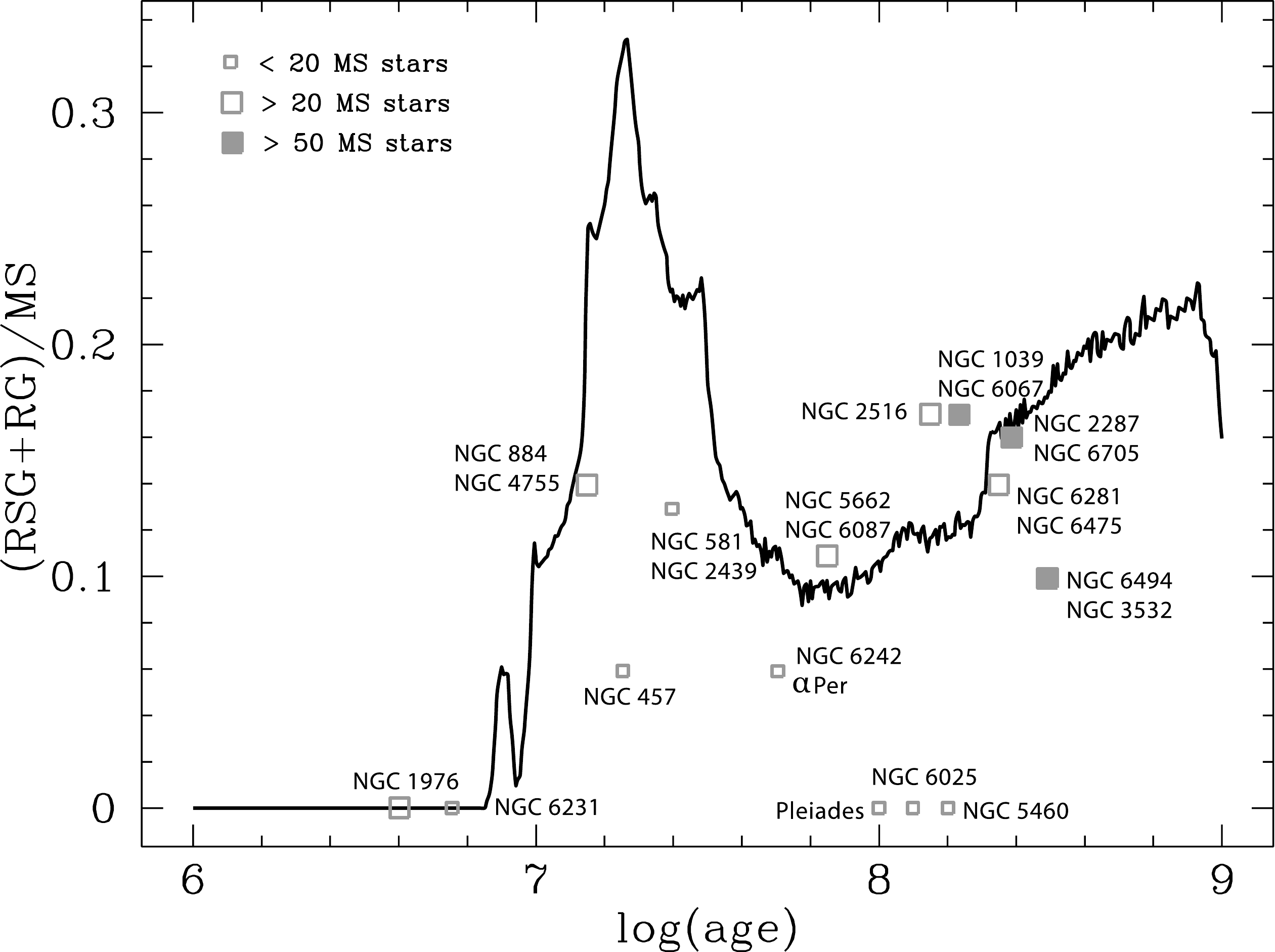}\hfill\includegraphics[width=.48\textwidth]{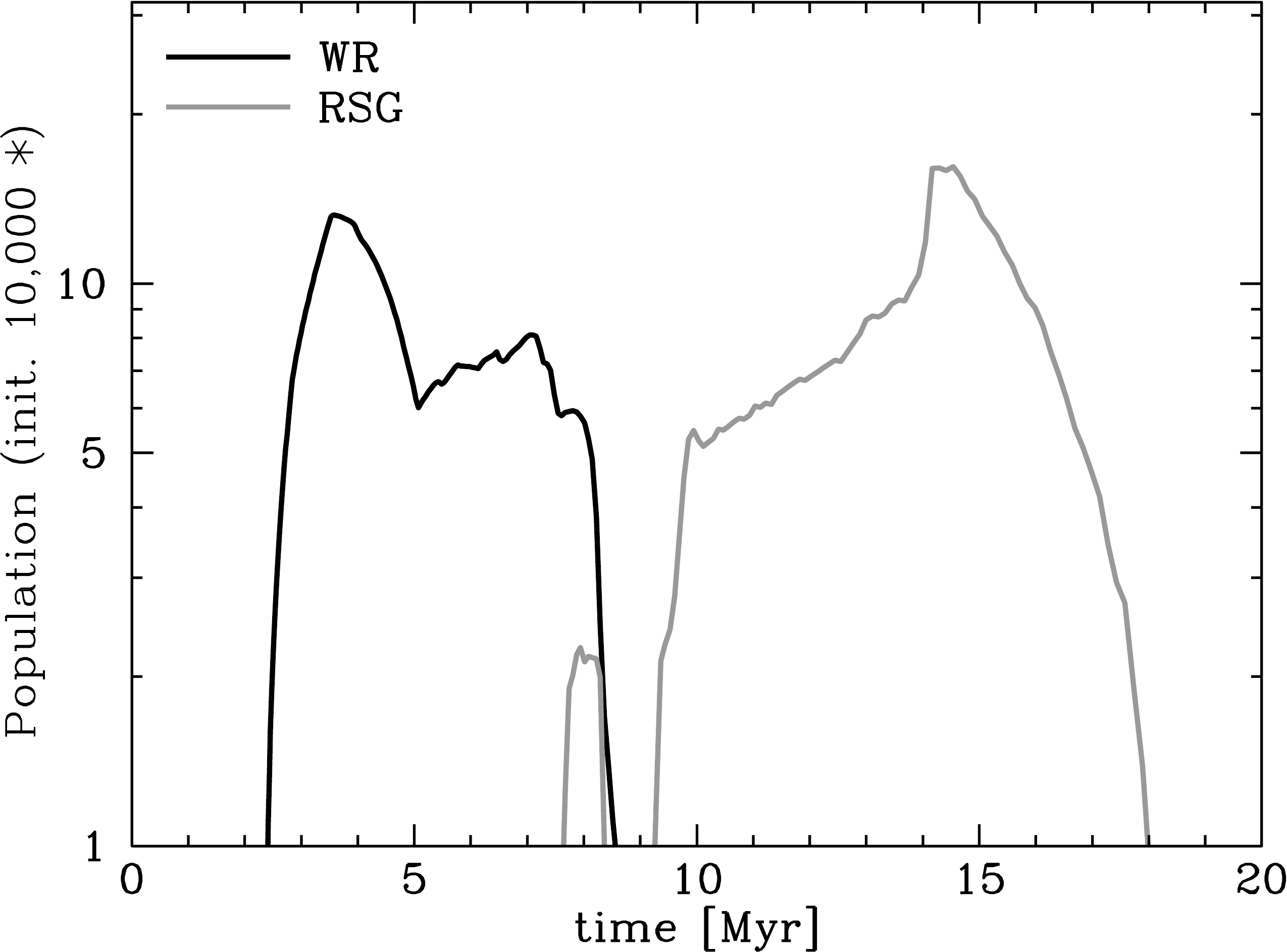}
\caption{{\em Left:} Evolution of the population ratio between red evolved stars and MS stars (in an interval of 2 mag below the turn-off) (theoretical curve: black solid line; observations in open clusters: grey squares, size and filling according to the MS star sampling). {\em Right:} Theoretical evolution of the number of WR (black line) and RSG (grey line) in a cluster of 10,000 stars at birth.\label{FigRSGPopCoEval}}
\end{figure}
%------------------------------------------------------------------------------------------------------------------------------------------------
\subsection{RSGs in coeval clusters}
The evolution of a population of red evolved stars in a cluster can be understood in the view of stellar evolution as follows (Fig.~\ref{FigRSGPopCoEval}, {\em left}):
\begin{itemize}
  \item at very young ages, no star has yet had time to become a RSG, so none is expected;
  \item around 10 Myr, one expects the number of RSGs to rise steeply, with a peak when high-$L$ RSGs (i.e. those arising from the most massive models going through a RSG phase) dominate the population;
  \item a drop-down occurs in the age range of wide-loop Cepheids;
  \item the number increases again when low-mass red giants start dominating.
\end{itemize}

In Fig.~\ref{FigRSGPopCoEval} ({\em left}), we evolved an initial population of 10,000 stars and counted as RSG the $M_\text{ini}\geq9\,\msol$ stars having $\log(\teff)\leq 3.70$ and as MS stars all non-evolved stars populating an interval of 2 magnitudes below the turn-off (TO). When we compare the theoretical curve to observed open clusters, we expect a large dispersion because we run into a stochasticity induced by small-number statistics. Typically, all clusters with a point falling well below the theoretical expectation contain less than 20 MS stars (2 magnitudes below the TO), the only exception being the NGC 3532 + NGC 6494 point.

Figure~\ref{FigRSGPopCoEval} ({\em right}) shows the evolution of WR-stars  and RSGs populations respectively. It is interesting to note that on theoretical grounds, we expect practically no overlap between both populations in the frame of single-stars evolution. Hence if some WRs would be observed in a cluster of RSG age (between around $10$ to $20$ Myr), it could provide a test for the binarity channel associated to the formation scenario of WR stars \cite[and references therein]{Eldridge2008a}. The opposite case (RSGs in clusters of WR age, {\em i.e.}, younger than $8$ Myr) would rather challenge the homogeneity of the population (cluster is not coeval, chance alignment, \dots).

%------------------------------------------------------------------------------------------------------------------------------------------------
\subsection{RSGs in constant star-formation regions}
The study of general populations is often done throughout a range of metallicities, and thus of different galaxies. In this case, the RSGs counts are performed in a restricted luminosity range ($\log (L/\lsol) \geq 4.9$), in order to avoid a contamination by red giants. We thus adopted the same criterion, using the Ekstr\"om \etal\ \cite*{Ekstrom2012a} grids and similar grids at $Z=0.002$ for which publication is in preparation.

\begin{figure}
\includegraphics[width=.55\textwidth]{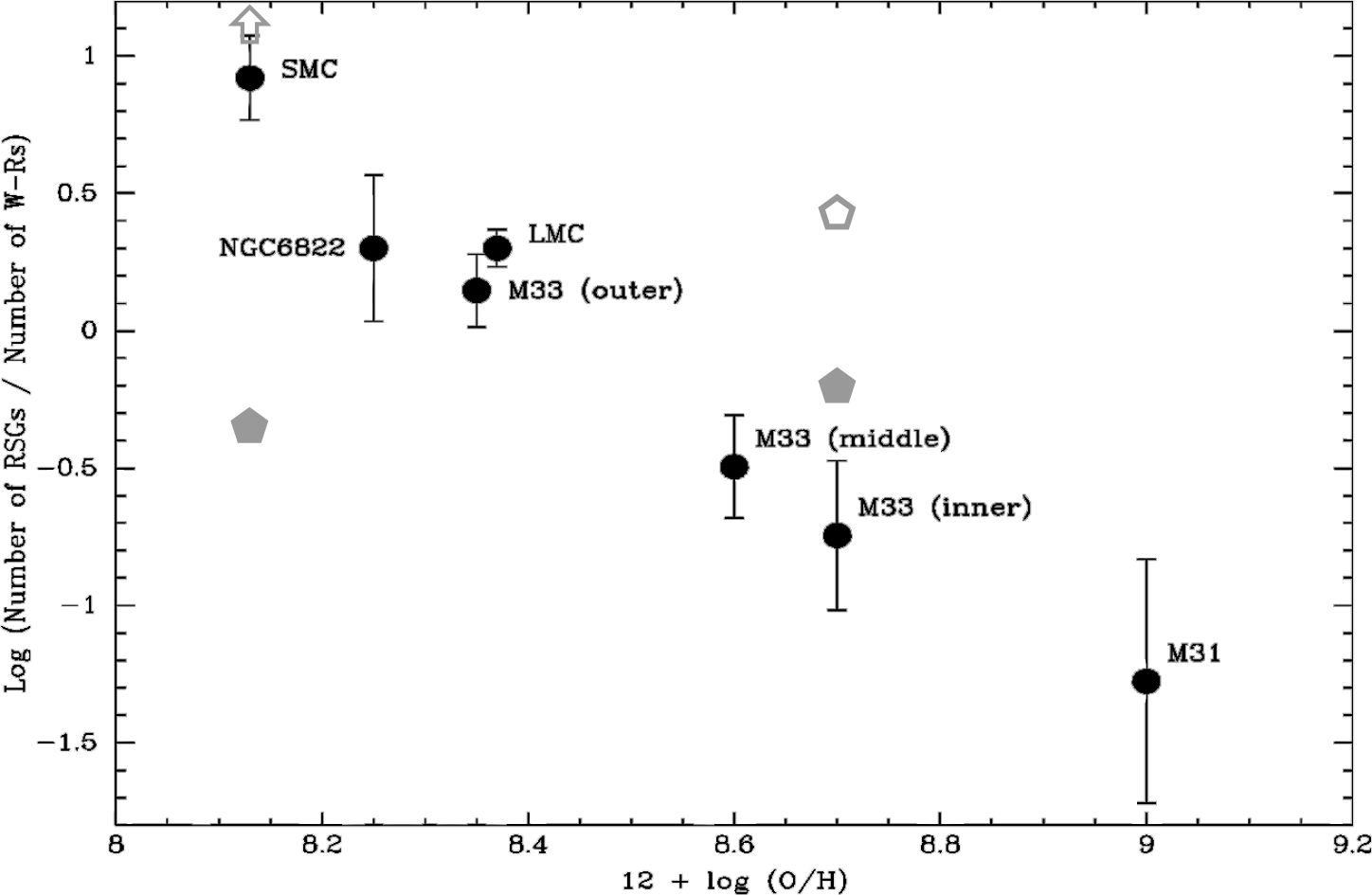}\hfill\includegraphics[width=.45\textwidth]{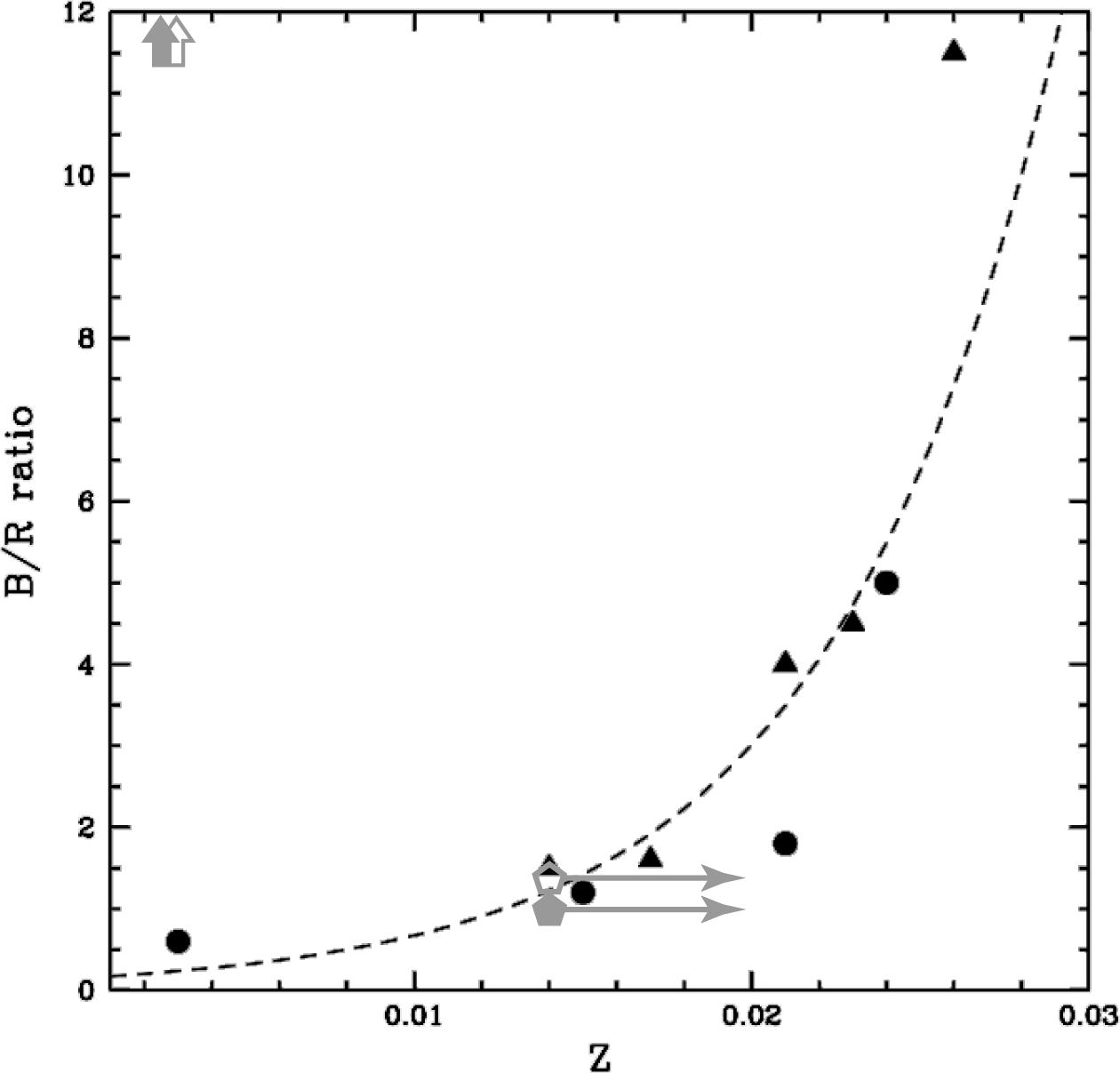}
\caption{Population-ratio predictions from the Ekstr\"om \etal\ (2012) models (grey pentagons: non-rotating, open symbol; rotating, filled symbol) overplotted on observational results from the literature. Note that the figures are exactly as published: when a theoretical point falls beyond the boundary of the plot, it is replaced by an arrow. {\em Left:} RSG to WR ratio. Figure~11 of Massey (2002). {\em Right:} blue- to red-supergiant ratio. Figure~1 of Eggenberger \etal\ (2002, black triangles: '$B$' contains O-, B-, and A-type stars; black circles: '$B$' contains only B-type stars).\label{FigRSGPopCstSFR}}
\end{figure}

Massey \cite*{Massey2002a} reports the values of the RSG/WR ratio observed in the metallicity range going from M31 ($12+\log(\text{O/H})=9.0$) down to the SMC ($12+\log(\text{O/H})=8.13$). He finds a linear increase of this ratio with lowering the metallicity (see Fig.~\ref{FigRSGPopCstSFR}, {\em left}). When we compare these results to models, there is a problem. Without rotation, the predicted ratios are much too high (by 0.5 to 1 dex), but the metallicity trend is well reproduced. With rotation on the contrary, the solar value is in relative agreement with the observations, but the metallicity trend is not reproduced and the SMC value is much too low. Note that we used the same initial velocity ($\vvc=0.40$) as average rotation rate than for the \zsol\ grids, and it might be too rapid for the average population at $Z=0.002$. A slower rotation rate might better reproduce both the values and the $Z$-trend.

Looking at the BSG/RSG ratio as reported by Eggenberger \etal\ \cite*{Eggenberger2002a} (Fig.~\ref{FigRSGPopCstSFR}, {\em right}), the situation is not better. While at solar metallicity, the values of both rotating and non-rotating models are quite in agreement with the observations, we definitively lack RSGs at low metallicity. The values for the SMC are much too high (94 for the non-rotating models, and 87 for the rotating ones, beyond the figure's boundary). This reflects the late crossing of the HR diagram at low $Z$. Note that we saw in Sect.~\ref{SecRSGCrossing} that varying the rotation rate had various outcomes on the time of the crossing. Based on duration considerations, we find for the $15\,\msol$ rotating at $\vvc=0.40$ a BSG/RSG ratio of 65, while if we consider a velocity distribution as proposed by Huang \etal\ \cite*{Huang2010a}, the same ratio drops down to 16, still too high but not as dramatically different. It points nonetheless to a possible solution to the population-ratio problem: by having the stars crossing the HR gap much earlier after the MS, which would increase the duration of the RSG phase.

%===================================================================================
\section{Conclusion}

The RSG phase is a delicate one for theoretical stellar models, because stars in this phase are dominated by mechanisms that are a challenge for 1D computations: convection and mass loss. Improvements in the observations of the RSGs mass-loss rates are highly needed to get reliable prescriptions to include in the evolution codes. Developments brought by 3D modelling of convection are becoming available and might bring physical prescriptions that can be included in the 1D codes.

The conditions for a star to become a RSG are sensitive to the structure at the end of the MS, mainly the size of the core and the amplitude of its contraction after H exhaustion. It is modified by overshoot, rotation, metallicity. Comparison of stellar-population ratios brings an observational constraint that points to a lack of RSGs produced at low metallicity by the models.

Let us conclude by quoting Langer \& Maeder \cite*{Langer1995a}, who finish their paper entitled precisely \textbf{\em The problem of the blue-to-red supergiant ratio in galaxies} with these words: {\em "We have to conclude that massive star models in the considered mass range still lack some significant physical ingredient. However we want to emphasise that this does not imply that the results of massive star theory have to be questioned altogether. The B/R-ratio is a quantity which is known to depend extremely sensitive on the model parameters. It is thus a welcome amplifier which can (and finally will) be very useful to constrain the model physics very accurately."}

We see that nearly 20 years later, the improvements brought to the models are not yet able to solve this longstanding problem. On the point of view of the physics, we understand the behaviour of our models, but Nature seems not to be eager to follow these rules. Since Nature is always right, we need to explore further the challenge she addresses.

%%-----------------------------
%%      your bibliography
%%-----------------------------
\bibliographystyle{astron}
\bibliography{BibTexRefs}

\end{document}